# Gross patient error detection via cine transmission dosimetry

Nguyen Phuong Dang[(1)], Victor Gabriel Leandro Alves[(1)], Mahmoud Ahmed[(2)] and Jeffrey Siebers[(1)]
[(1)]University of Virginia Health System, Charlottesville, VA
[(2)]Vanderbilt University Medical Center, Nashville, TN

**Purpose:** To quantify the effectiveness of EPID-based cine transmission dosimetry to detect gross patient anatomic errors.

**Method and Materials:** EPID image frames resulting from fluence transmitted through multiple patients anatomies are simulated for ~100 msec delivery intervals for hypothetical 6 MV VMAT deliveries. Frames simulated through 10 head-and-neck CTs and 19 prostate CTs with and without 1-3 mm shift and 1-3 degree rotations were used to quantify expected in-tolerance clinical setup variations. Per-frame analysis methods to determine if simulated gross errors of (a) 10-20 mm patient miss alignment offsets and (b) 15-20 degree patient rotations could be reliably distinguished from the above baseline variations. For the prostate image sets, frames simulated through the reference CT are intercompared with (c) frames through 8-13 different CT's for the same patient to quantify expected inter-treatment frame variation. ROC analysis of per-frame error discrimination based upon (i) frame image differences, (ii) frame histogram comparisons, (iii) image feature matching, and (iv) image distance were used to quantify error detectability.

**Results:** Each error detection method was able to distinguish gross patient miss-alignment and gross rotations from in-tolerance levels for both H&N and prostate datasets. The image distance algorithm is the best method based on AUC.

**Conclusion:** In-field gross error detection was possible for gross patient miss-alignments and incorrect patients. For prostate cases, the methods used were able to distinguish different patients from daily patient variations.

**Keywords:** radiation therapy, real-time, EPID, QA, ROC analysis

# 1  INTRODUCTION

Achieving the appropriate patient dose in radiation therapy requires delivery of the intended beams and that the correct patient is appropriately aligned with respect to those beams. While significant effort has gone towards verifying the former, in the form of IMRT pre-treatment delivery quality assurance (DQA), the latter is assumed to be achieved via patient pre-treatment alignment protocols, and is in some cases verified to be unchanged during the treatment delivery by use of e.g. radiofrequency (RF) Calypso markers [1], surface imaging [2-4], intermittent during treatment image [5,6], etc. Relatively little has been published on utilizing the patient radiation treatment exit fluence to verify that the correctness of the patient pose.

The use of electronic portal imaging device (EPID) in DQA and patient dosimetry is becoming commonplace due to its convenience of setup, two-dimensional measurement and high spatial resolution [7-11]. The EPID can record images produced by the exit fluence after going through a patient from an incident beam. This exit image can then be used to verify patient treatment, either by comparing to the predicted exit image [12-15] or by reconstructing the dose delivered to the patient [16-18]. These are powerful methods to detect errors which occur during a patient's treatment delivery. A recent study, which evaluated incident reports gathered over a 2-year period, shows that EPID-based dosimetry *in vi*vo first fraction measurements have better performance in detecting incidents compared to software-based error detection system (rules-based and Bayesian network checks), especially with incidents related to patient positioning [19]. In addition, the comparison of VMAT QA systems (Delta4, EPID-based dosimetry, and log file) using three possible scenarios (uncertainty, miss-calibration, and worst-case scenario) also shows that EPID-based dosimetry has the highest complex error detection sensitivity [20].

In this study, we develop and study a real-time patient-pose error detection module developed for inclusion in our robust real-time delivery verification QA based on Swiss cheese model error detection (SCED) method [21,22].  The SCED error detection model consists of a series of checks designed for real-time QA of cine EPID images. The goal of this study is to quantify the effectiveness of EPID-based cine transmission dosimetry to detect gross (dosimetrically consequential) patient anatomic errors in real time. Two types of patient errors are included in this work: (a) gross patient miss-alignments (patient shifting and rotation), and (b) incorrect patient (distinguishing gross patient miss-alignments or wrong-patient from in-tolerance inherent patient variations). Each error type was analyzed to determine whether or not it could be detected by use of per-frame EPID analysis.

# 2  METHODS AND MATERIALS

The real-time error detection method we propose tests for gross anatomy errors by comparing EPID image frames acquired during the treatment delivery with frames predicted through a recent (planning or more recent) patient CT image set.  For each frame, EPID in-aperture fluence (image) changes from expected are used to determine if the patient pose differed from the expected.  Figure 1 diagrams the EPID-based real-time treatment verification system for a dynamic IMRT/VMAT treatment. The predicted EPID images are calculated in advance from the treatment plan and the recent patient CT. The measured EPID images are acquired during treatment, in cine mode (frame-by-frame) at a ~10 Hz rate.  Therefore, an image frame corresponds with ~100 msec of beam delivery.  Image frames acquired during the treatment beam delivery are compared to the predicted ones.  The verification system performs the frame-by-frame agreement (pass/fail) decision based on image comparison metrics. In the final step, the system cumulates the individual decisions (all frames up to the current acquired frame) and makes a pass/fail single decision regarding the beam delivery.  By utilizing a series of single EPID frame image checks with per-frame images, the error detection task is reduced to simple comparison operations that can be performed in real time.

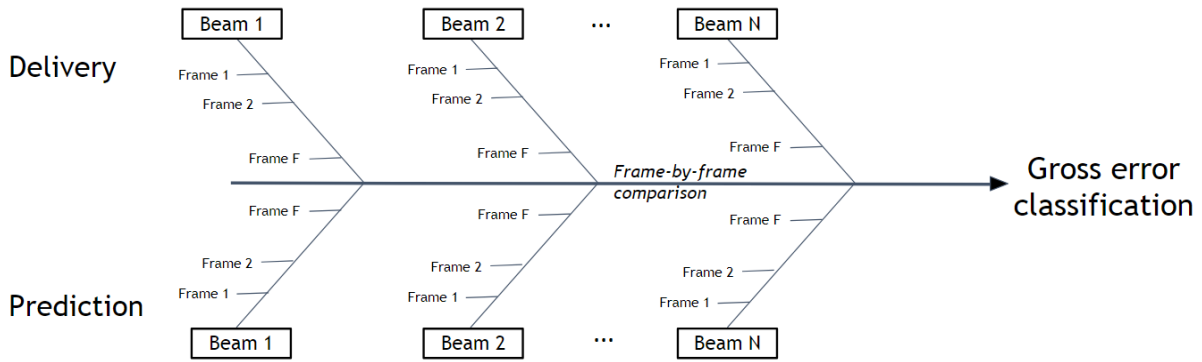

Figure 1. Diagram of the EPID-based real-time treatment verification system. The EPID frame images are acquired in cine transmission dosimetry at a ~10 Hz rate. For each beam, measured frames (Frame 1, Frame 2, …, Frame F) are acquired and compared to the corresponding frames predicted through the patient image. A frame-by-frame comparison metric is calculated to decide whether the two frames are in agreement. The verification system accumulates frame-by-frame decisions for overall error classification.

In practice, gross errors are detected by comparing the predicted EPID transmission image and the measured one.  To ensure a controlled investigation, in this study we inter-compare images predicted through CT image sets with and without simulated errors. That is, we perform a theoretical estimation of EPID-based gross error detection. This is necessary to ensure that the patient anatomies and associated perturbations are known. Furthermore, simulation is required to ensure that we have a sufficient number of true errors in the dataset to perform meaningful analysis.

## 2.1 EPID image generation

The frame image recorded by the EPID is predicted by DRR-like algorithm that ray-traces incident fluence through the CT image set, then convolves the exit fluence with Monte Carlo-based convolution kernels, similar to the kernels described in [23-25]. In the model, transmitted primary fluence incident on the EPID is obtained via the ray tracing, while patient-scatter fluence is approximated by convolving the primary transmitted fluence with convolution kernels specific to the patient thickness traversed. The prediction model reads in the DICOM treatment plan file and the patient CT data and calculates EPID images as a function of control points (fractional MU). For this work, we simulated 360° 6 MV VMAT fields from a TrueBeam linac (Varian Medical Systems, Palo Alto, CA, USA) equipped with a Millennium 120-leaf multileaf collimator (MLC). The EPID imager system contains an amorphous-silicon aS1200 flat panel detector placed at 150 cm source-to-detector-distance (SDD). The imager has 1280x1280 pixel resolution over the 43x43 cm$^2$ detector area. The incident fluence in our prediction model was optimized for our linac and EPID combination using a series of open fields (3x3 to 40x40

cm$^2$, no phantom in beam path) and validated against measured EPID images with different thickness slab phantoms (1-30 cm) in the beam path. The maximum discrepancy between predictions and measurements was less than 5%.

Cine EPID image frames resulting from treatment beam fluence transmitted through a patient's anatomy are simulated for 10 head-and-neck (H&N) former UVA patients and 19 prostate patients from the Netherlands's Cancer Institute (NKI), which includes multiple CT images per patient [26]. We designate the set of patients to be *{a, b, …}* with differing patient poses for treatment planning and treatment delivery, given by image sets *n={0, 1, …N}*. For simplicity in discussion, consider patient *a*. A treatment plan on image set *0* is designated *Pa(0)*. The EPID image frames predicted through image set *a(0)* is then *Ip(Pa(0), a(0), f)* where *f=1,2,…F* is the frame index. For ease in presentation, we drop the frame index in our nomenclature, so the set of EPID image frames becomes *Ip(Pa(0), a(0))*. The set of EPID images predicted for patient pose *a(1)* for plan *Pa(0)* is then *Ip(Pa(0), a(1))*.

## 2.2 Cine image comparisons for gross error classification

Patient variations for simulated treatments are monitored with the EPID by frame-by-frame comparisons of *Ip(Pa(n), a(n)))* with *Ip(Pa(n), a(m))* with *n≠m* (different image, same patient) and *Ip(Pa(n), a(n)))* with *Ip(Pa(0), b(n))* (different patient).

The following scenarios were simulated to analyze the system's error detectability:
(a) Patient miss-alignment simulations: CT image sets are shifted and/or rotated with respect to the treatment beam(s) resulting in miss-aligned image sets (e.g. *a(i)=T(a(0)*, a translated and/or rotated image set. Baseline in-tolerance miss-alignments of 1-3 mm shifts and 1-3° rotations were compared with gross 10-20 mm shifts and 10-15° rotations to determine the ability to discern gross miss-alignments from baseline.
(b) Patient pose variations: EPID image sets consist of predictions through alternative CT images from the same patient as the CT used to create the treatment plan, e.g. *Ip(Pa(n), a(m)), n≠m* and image sets using a different patient's image set, e.g. *Ip(Pa(0),b(n))*. For *Ip(Pa(n), a(m)), n≠m*, the variation is considered in-tolerance if the CTV or OAR dose difference between images *n* and *m* is <5%, and a gross error otherwise. A different patient is always considered a gross error.

This study classifies the detected gross errors into miss-alignment and incorrect patient errors. For each type of gross error, the EPID image prediction system was used to simulate "error" and "baseline" deliveries for testing the error detection algorithm. The error detection algorithm was evaluated for detection reliability using this simulated dataset. Figure 2 shows the diagram of frame-by-frame gross error classification. Predicted images are created for each simulated error mode. A given reference patient pose, e.g. *a(0)*, is perturbed / offset multiple times to generate different *a(n)* test poses. Different predicted EPID image sets are used to assess different types of gross errors. A decision tree classification is utilized for different errors/variations observed.

For the miss-alignment type of errors, frames simulated through 10 head-and-neck CTs and 10 out-of-19 prostate CTs (the rest were used for validation as in Sec. 2.3) were used to set the frame error threshold. The detection tolerance is set by comparing baseline and gross error image predictions with the treatment plan-based images.

For incorrect patient type of errors, we utilize 10 out-of-19 prostate cases since these cases have 8-13 CTs acquired throughout the treatment course. For each patient, we use the first-fraction prediction as a treatment planning reference. The incorrect patient error is studied by comparing image predictions through a CT set of a different patient (e.g. *b(n)*), with the treat intended (*a(0)*) based predictions.

Frame-by-frame comparison metrics were used to determine whether a given EPID predicted image was in agreement with the initial treatment plan. In this study, we utilize several comparison metrics which are
  (a) image difference, which is simply calculated by subtracting two images. Ideally, any non-zero difference indicates an error, however, this criteria could cause 100% error reporting rates due to in-tolerance patient variations and EPID output statistical variations. Therefore, a threshold is used to differentiate different (higher than threshold) from similar (below threshold) images.
  (b) intensity histogram comparison, by comparing histograms of pixel intensity values from the images. The histogram is a 1-D plot showing the number of pixels from an image at each different intensity value. Histogram comparison is likely to be good at identifying an incorrect patient error since it should detect changes in beam attenuation through a patient.
  (c) image feature matching, by utilizing feature extraction techniques and comparing the extracted features from images. In this work, image peak features were extracted by using *scikit-image* tool, an image processing library in Python language [27]. The minimal distance between two feature sets was calculated after matching (via rotation and translation). Feature matching is likely to be good at finding offsets of patient heterogeneities that exist within the field.
  (d) image distance, this metric is usually used to measure the similarity between two images. The image comparison can be made via widely applicable distances such as Manhattan distance, Euclidean distance, Hausdorff distance, etc [28]. In general, there is no single distance metric that can work well for all tasks, each specific task requires a different property measurement. Therefore, we investigate the performance of each metric to determine which distance metric provides the best performance.

To determine optimal comparison metrics thresholds, the gross error detection algorithm was performed using a tree decision model with large sets of EPID image frames from the prediction system. We utilized receiver operator characteristic (ROC) techniques [29-31] to quantify the optimal metrics threshold of error detection algorithms. Within the ROC analysis, several thresholds will be used to classify EPID frame images to gross error and baseline categories.

In clinical practice, if error happens, a single decision regarding gross error detection must be made during a patient treatment session. To account for this, we identify gross errors when the number of failing EPID frames (failed-frame rate) exceeds a given percentage of the total frames in the delivery (as in Figure 2). This percentage should be chosen to be able to detect gross errors with low false positive error detection rates.

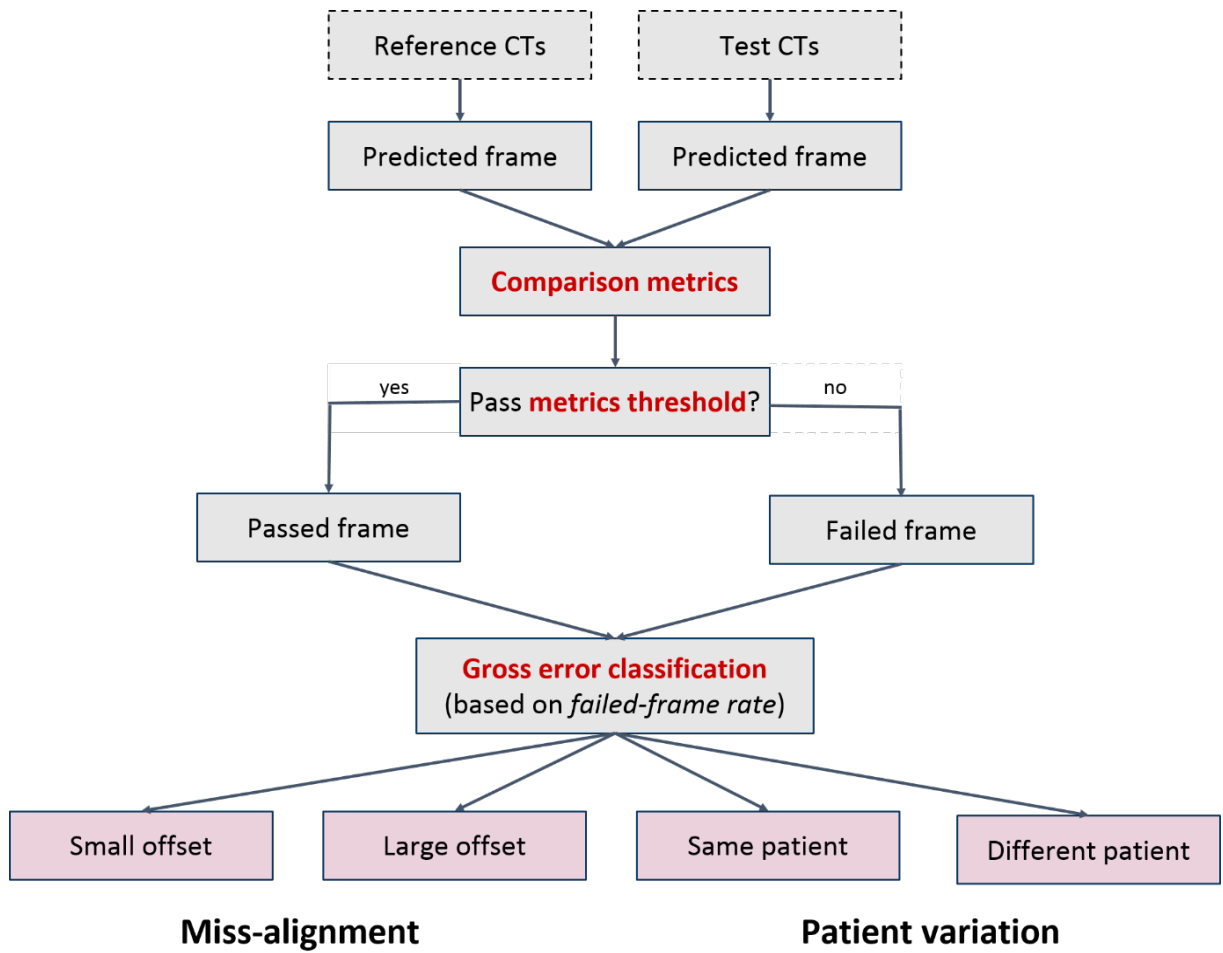

*Figure 2. Flow chart for real-time gross error classification. For miss-alignment error classification, the baselines are 1-3 mm shifts and 1-3° rotations, and the gross errors tested are 10-20 mm shifts and 10-15° rotations. For patient variation error classification, the baselines are the same patients with different treatment daily changes, and the gross errors are different patients.*

## 2.3 Testing cases

To validate the ability of our gross error detection algorithm, additional gross error test cases were simulated with the same treatment plan discussed above, and the same error detection method was used as for the real-time dose verification. The 9 out-of-19 prostate cases from the NKI dataset were used in this validation. Each prostate case has 8-13 CTs made during the treatment course, which correspond to 8-13 treatment days. These CTs were used to simulate EPID frame images from daily patient changing variations. The frame-by-frame comparison metrics threshold and failed-frame rate derived from the previous section were used to make error detection decisions for each single day of treatment.

## 3 RESULTS

In this study, 231 treatment fields from 29 treatment plans (10 H&N and 19 prostate cases) were simulated and analyzed. Predicted EPID images from 10 H&N and 10 prostate treatment plans were used to establish thresholds for the frame-by-frame image comparison method; and 9 prostate treatment cases were used to validate this method. The optimal metric for frame-by-frame comparisons was found in Sec.3.1 and then applying to per-beam error classification in Sec. 3.2.

## 3.1 Frame-by-frame comparison metrics

Among the frame-by-frame image comparison metrics listed in Sec. 2.2, the image distance metric has several options. Our goal is to discriminate gross patient errors (large image miss-alignment and incorrect patient) from the baseline (minor image offset, different image of correct patient). Figure 3 shows the histograms from the same patient, different image set (baseline) and different patient (gross error) using different image distance metrics (Manhattan, Euclidean, Canberra, and Hausdorff) for a single prostate case. The most discriminative distance metric is the one which has the smallest overlapped area between baseline and gross error histograms. For this test case, the Manhattan distance provides the least overlap, therefore, we use the Manhattan distance as the image distance metric for further frame-by-frame comparisons.

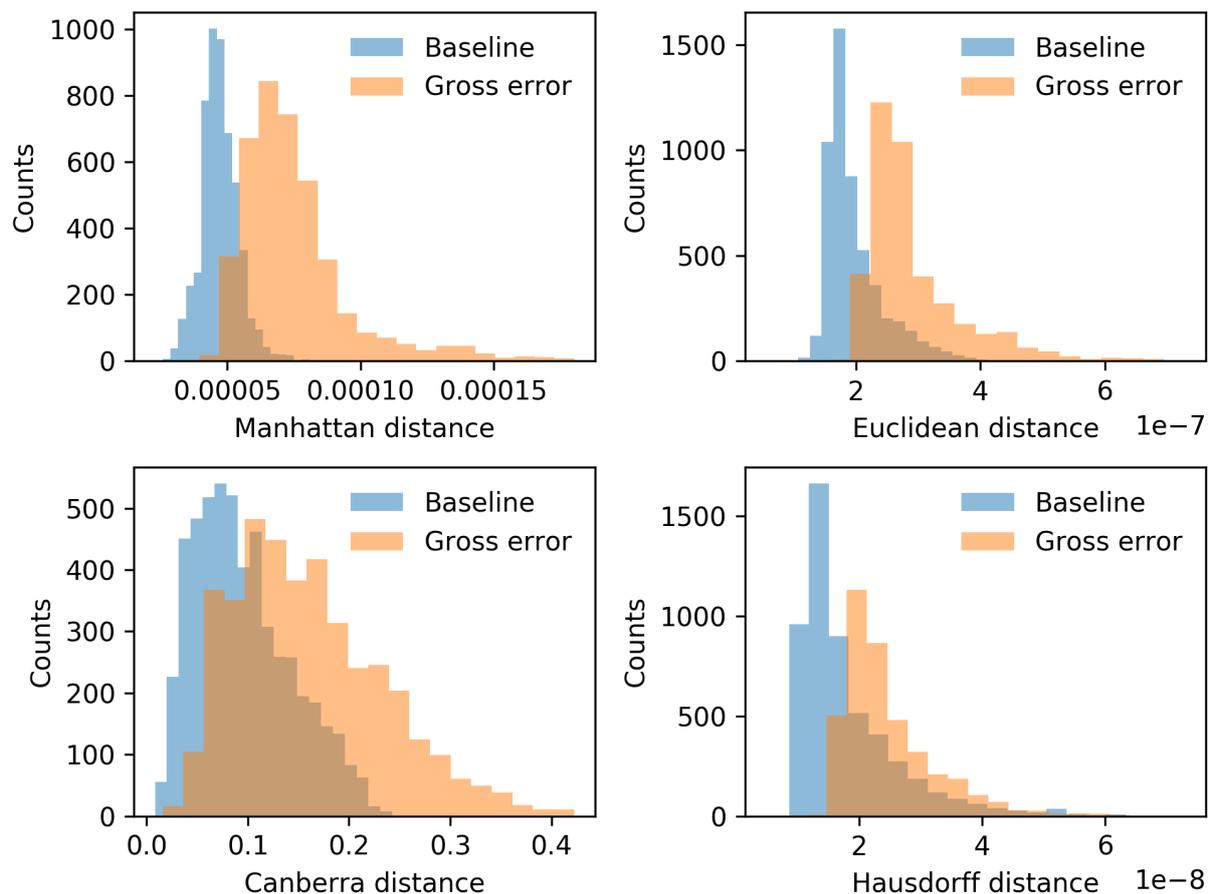

*Figure 3. Comparison between baseline and gross error histograms using different image distance metrics for a single prostate case. The Manhattan distance metric (top left) gives the best discriminating power compared to Euclidean (top right), Canberra (bottom left), and Hausdorff (bottom right) distance metrics.*

The mean ROC curves for the frame-by-frame analysis, where each frame is considered a sample from 10 H&N and 10 prostate patient datasets are shown in Figure 4. Four different comparison metrics were used as described in Sec. 2.2, each metric, excepting feature matching, was normalized by dividing the number of active pixels. Active pixels are defined as pixels which have intensity values 3 times larger than the average noise values of a measured dark field. Figure 4a indicates the extent that large patient shifts and rotations can be discerned from typical in-tolerance variations observed on a fraction-to-fraction basis. The areas under ROC curves (AUCs) were 0.71 (image difference), 0.80 (histogram comparison), 0.64 (feature matching), and 0.88 (image distance). Figure 4b shows the ROC curves for distinguishing an incorrect patient from the same patient with inter-fraction variations on a daily basis. For this error detection, we only utilized the 10 prostate datasets since these datasets have CTs taken on (8-13) different days. The AUCs were 0.56 (image difference), 0.82 (histogram comparison), 0.61 (feature matching), and 0.91 (image distance). Both figures show poor performance for other methods compared to the image distance one. The optimal operating point for gross error classification threshold is determined as the point which has the highest TPR/FPR ratio. In practice, the closest point to point (0, 1) on the ROC curve tends to have the best performance (highest TPR and lowest FPR combination). Therefore, the optimal thresholds are chosen to have the smallest distance to point (0, 1). The average of miss-alignment and incorrect patient optimal thresholds was taken to be the single frame-by-frame gross error classification; a calculated frame image comparison metric below this threshold is considered as a failed frame.

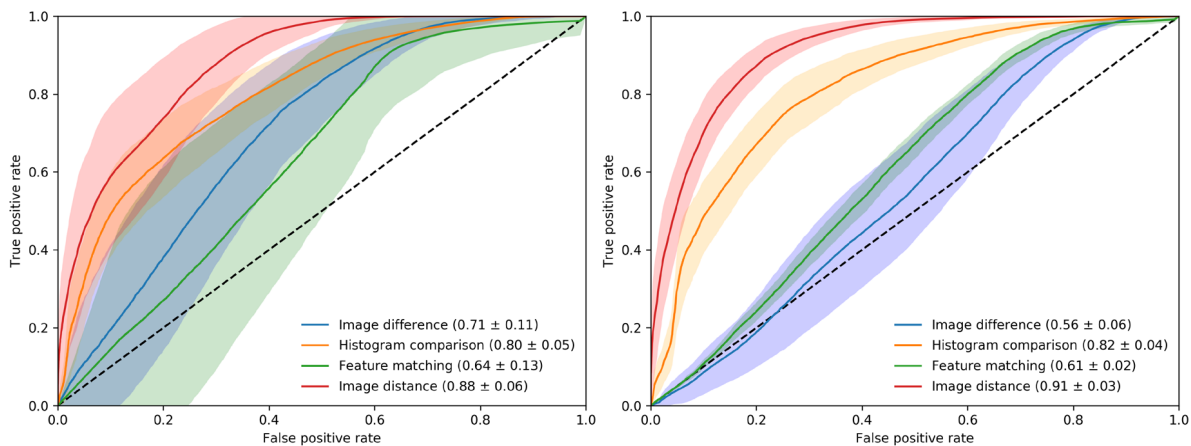

Figure 4. Mean ROC curves for (a) miss-alignment classification in left plot, and (b) incorrect patient in right plot; with different frame-by-frame comparison metrics. The solid lines show mean values, and shaded bands show standard deviations of ROC values calculated from 10 H&N and 10 prostate patient datasets. The numbers in brackets show AUC corresponding to each comparison metric.

## 3.2 Per-beam error classification

The frame-by-frame comparison metric and cut-off threshold identified in Sec. 3.1 performs gross error classification of a single EPID frame. However, a single frame classification does not provide sufficient accuracy for clinical use. For example, with a 10% FPR classification as in Figure 4, a baseline setup will give frame-by-frame comparison failure rate of about 10% while the corresponding error setup will give failure rate of about 70%. With ~1000 frames per patient, a 10% per-frame FPR ensures that each patient would be incorrectly labeled as being in error. Therefore, instead of using error classification based on detecting single failure frame, we combine classification over frames acquired during treatment and construct an error classification based on the failure rate of the frames.

The verification system can identify the gross error scenarios based on the fact that each scenario has a different size of patient variation and, as a consequence, have different frame failure rates. In Figure 5, the failure rates of frames (failed-frame rates) from patient variation baselines and gross errors are shown. The patient miss-alignment classification threshold was used to calculate these rates with attempting of merging two frame-by-frame classification into a single per-beam/per-fraction classification. Miss-alignment categorized failed-frame rates were shown by filled areas and incorrect patient ones were shown by solid lines. Patient daily variations in setup (red line) are larger than our estimated baseline offset tolerances (1-3 mm shifts and 1-3° rotations) but still within our same patient variations (10-20 mm shifts and 10-15° rotations). The incorrect patient gross error always has high failed-frame rates (above 90%).

For clinical gross error detection purposes, we propose 3 variation levels which can be used to indicate the how large of patient variation during radiation therapy treatment: (i) first level indicates in-tolerance variation, (ii) second level indicates out-of-tolerance (large) variation, and (iii) third level indicates very large patient variation or wrong patient. The rate thresholds are chosen in order to distinguish different types of gross patient errors. From Figure 5, the proposed thresholds for these levels are 30% between first and second levels and 80% between second and third levels.

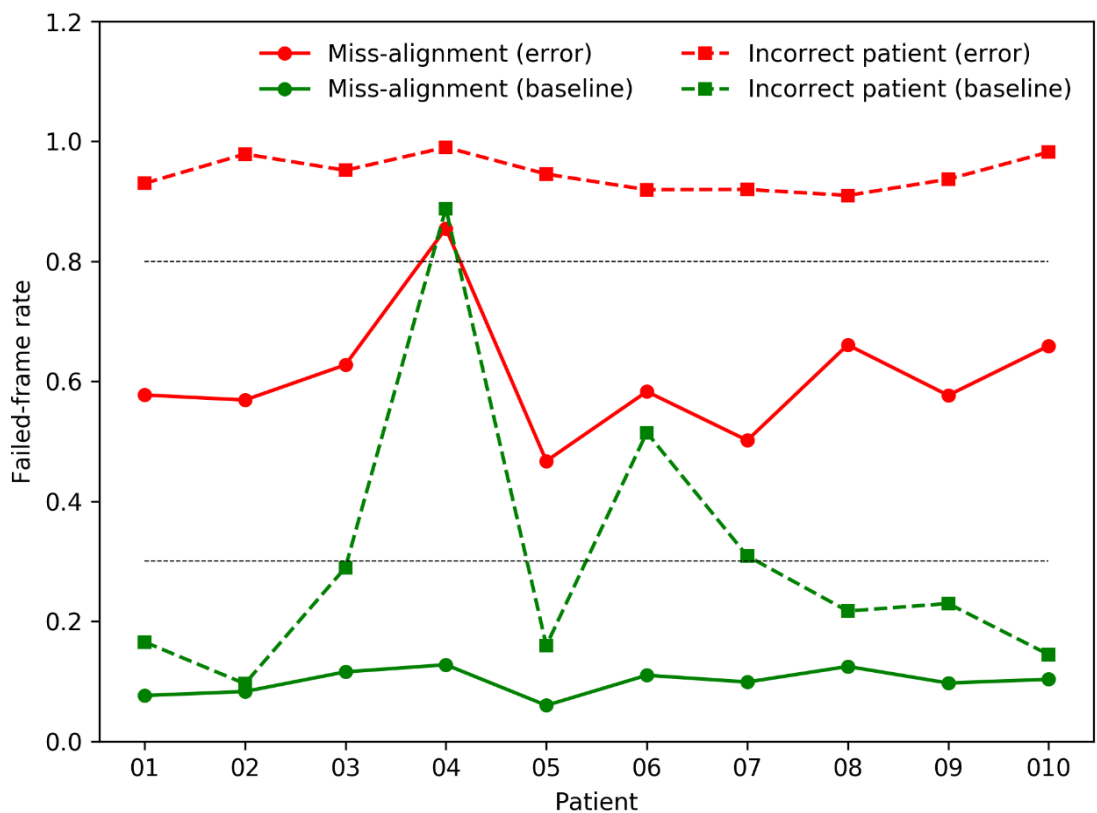

*Figure 5. Frame failure rates (failed-frame rates) from 10 prostate patient datasets. Solid lines show failed-frame rates from miss-alignment classification and dashed lines show failed-frame rates from incorrect patient classification. The green color indicates failure rates from baseline variations and the red one indicates failure rates from gross error variations. The black dashed lines (at fail-frame rates 30% and 80%) show the proposed tolerance levels for gross patient error detection determined from analysis of these 10 patients.*

## 3.3 Testing cases

For testing the ability of our gross patient error detection algorithm, the 9 NKI prostate cases that were not part of the initial test set were used. An automated treatment plan was created for each treatment day CT dataset, then EPID frame images were calculated for treatment plan and CT dataset combination. Frame error detection rate (failure rate) was calculated by comparing frames from the first treatment day with the ones from the following treatment days as described in Sec. 3.2. Figure 6 shows the comparison of error detection rates of 9 NKI patients from all combinations of treatment plans and CT datasets. Most of the patients have error detection rates at in-tolerance variation level (< 30%), three patients (numbers 03, 06 and 09) at out-of-tolerance variation level (< 80%), and one patient (number 02) has very high error detection rates (more than 80%). It is notice that the high error detection rates of Patient 09 can come from one specific treatment plan or acquired CT dataset.

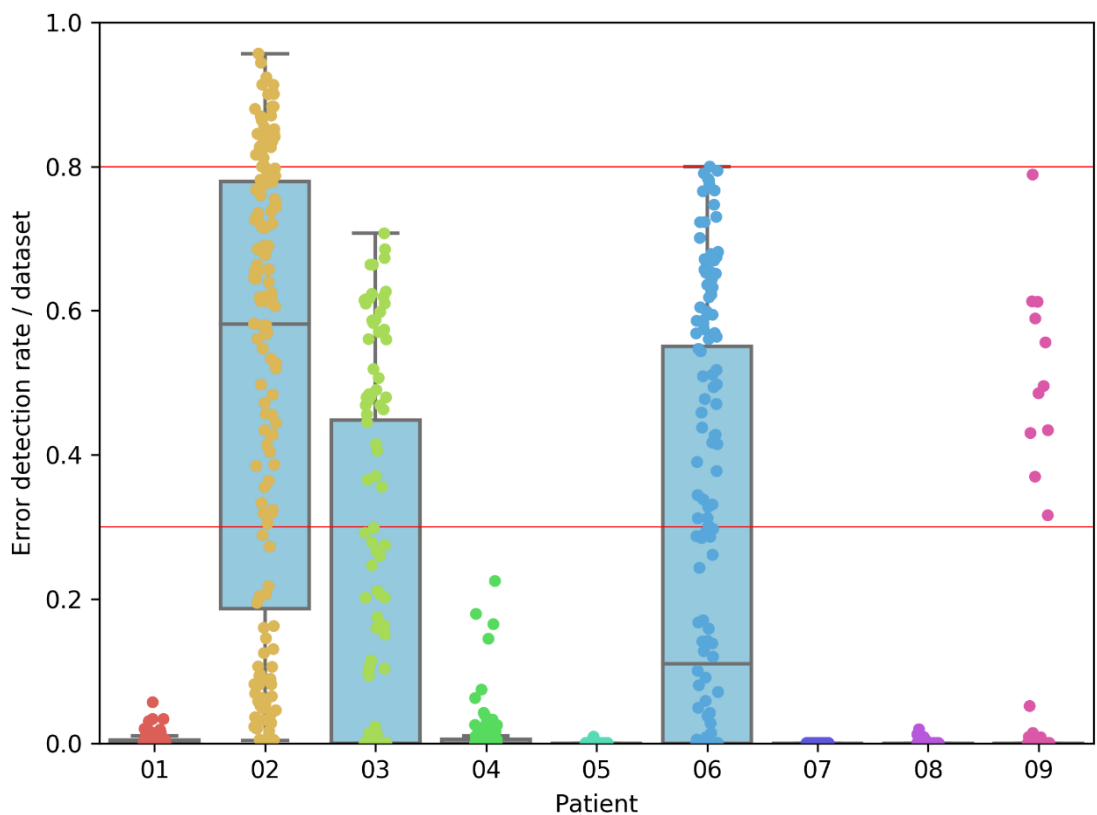

*Figure 6. Comparison of error detection rates of 9 NKI prostate treatment patients. The detection rates were calculated from all combinations of treatment plans and datasets. The red lines indicate proposed tolerance levels for gross patient error detection as described in Sec. 3.2.*

Figure 7 shows more detail about the error detection rates from two patients 02 (top row) and 09 (bottom row). The left plots showing correlation matrices between treatment plans and CT images indicate that our error detection system is robust against CT image variations. The error detection rates are consistent across all treatment plans as shown in the right plots. The bottom right plot in Figure 7 also confirms that the high error detection rates of Patient 09 seen in Figure 6 are from one specific acquired CT image set (dataset 4).

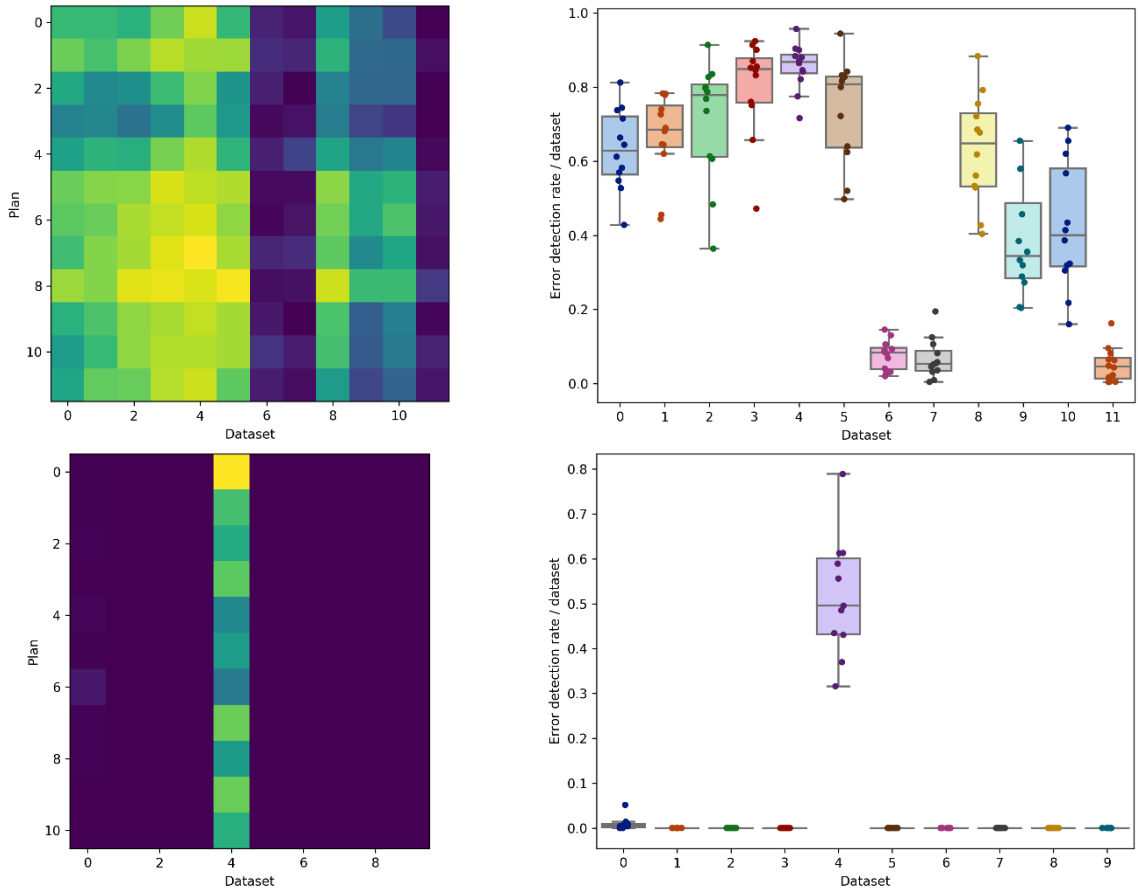

*Figure 7. Gross error detection rates from all combinations of treatment plans and datasets for Patients 02 (top row) and 09 (bottom row). Left: correlation matrix plots between treatment plans and datasets. Right: error detection rate distributions for each patient dataset.*

Figure 8 shows the prostate CT images from different datasets of Patients 02 (top row) and 09 (bottom row). The images from reference dataset were compared with the ones from error and non-error detection datasets. Large air bubbles in recta can be seen in both error detection datasets (middle column), which confirm that the error detections of these two patients are triggered by CT image variations.

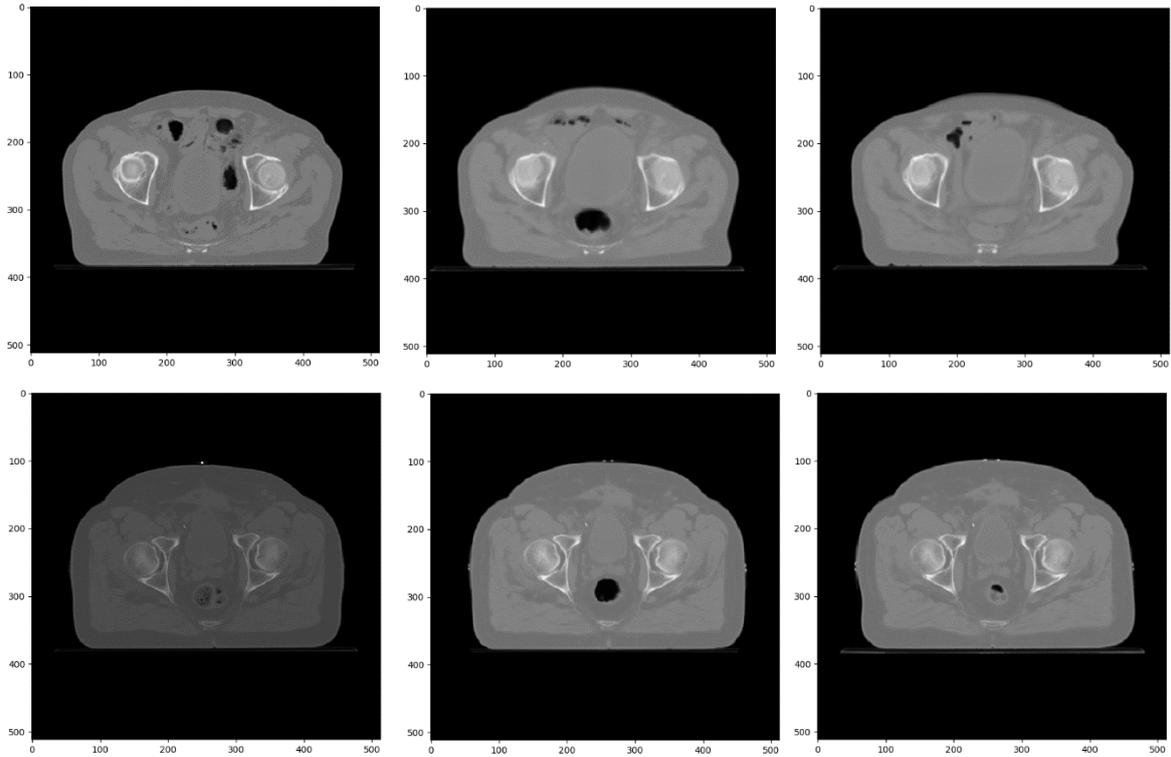

*Figure 8. CT images from Patients 02 (top row) and 09 (bottom row). From left to right: reference (1$^{st}$ day of treatment), error (dataset 4) and non-error (dataset 7 for Patient 02 and dataset 9 for Patient 09) detection datasets.*

## 4 DISCUSSION

The common errors in IMRT and VMAT can be categorized in two types: gross errors such as patient miss-alignment, incorrect collimator rotation, incorrect isocenter position [32,33], and non-gross errors such as CT data error or subtle treatment plan difference [33,34]. Current systems do not account for several error pathways and are unreliable in detecting the errors they were designed to prevent. Pre-treatment QA is impractical for on-line adaptive radiation therapy and does not monitor inter-fractional delivery variations (e.g. due to unintentional modification of delivery parameters or equipment malfunction) [35,36]. DQA and during-treatment linac performance monitoring (of e.g. MLC and output) is oblivious to the patient, including gross errors such as the wrong patient or inter- and intra-fractional patient changes that can negatively affect the patient's dose and outcome. Therefore verification of the actual treatment delivery as an end-to-end treatment check can reduce the likelihood that mistreatments occur.

This work has demonstrated the feasibility of EPID frame-based patient verification during radiation treatment using continuous comparison of EPID image frames (cine-mode) with predicted EPID image frames. This study also developed a method to quickly identify gross patient errors during radiation delivery. An advantage of this implementation is that real-time image acquisition can be used, so that the gross errors can be detected during treatment and therefore, prevent radiation under or overdosed for patients.

The selection of image distance metric for real-time treatment verification system was acquired from the study using ROC curves, however, it is just one of many possibility frame comparison metrics. Multiple metrics can be stacked together to give a better error detection ability since each of them may have discriminating power on certain types of error detection. This study already shows us distance metric has very high discrimination (AUC ~0.9), therefore, we won't have a large gain by using a combined metric in this situation. However, in clinical practice, the distance metric may not have good performance as in this theoretical calculation, and the combined metric can be considered as a better solution.

The optimal frame-by-frame comparison working points are chosen based on the optimal TPR/FPR ratio for 10 H&N and 10 prostate patients. There is potential to change these points and improve the system sensitivity or lower FPR; however, this would affect the gross patient error detection ability during treatment. The gross patient error variations (10-20 mm shifts and 10-15° rotations) was chosen to cover all patient daily variations in reality, as shown in Figure 5, where the red line represents the variation between different treatment day is within the blue area which represents the large variation of the same patient. This choice give us an opportunity to merge both gross patient error classifications (miss-alignment and incorrect patient) into one combined classification since the error setup of miss-alignment error is baseline setup of incorrect patient error.

On the other hand, by translating frame-by-frame classification based on distance metric into per-beam classification based on failed-frame rate, we can eliminate the incorrect labelling due to high FPR (~10%) in frame-by-frame classification by utilizing a large number of acquired frames. The large difference in failure rates between baseline and error patient setups helps us in establishing a threshold to separate them. A proposed 20-30% failed-frame rate threshold can eliminate the chance of incorrectly labelling a patient as being in error since the average FPR is about 10% which means the average failed-frame rate of non-error patient setup is only about 10%.

The method sensitivity has been tested for gross patient errors. Figure 6 and Figure 7 illustrated that the method is able to detect gross patient error by utilizing 9 NKI prostate treatment cases. The results show that the system can detect gross errors during treatment session. Once the system detects an error, the decision as to whether to stop the treatment and investigate the error would lie with radiation therapists who are in charge of the treatment. Figure 9 demonstrates the daily gross patient error detection with using 9 NKI prostate treatment patients as an example. Treatment days without any errors detected (error detection rate < 30%) are labeled in green; treatment days with high risk of error (error detection rate between 30% and 80%) are labeled in orange; and treatment days with errors (error detection rate > 80%) are labeled in red. With the results obtained in Figure 6 and Figure 7, we would expect that there are two treatment days with errors for Patient 02 and few days with high risk of errors for Patient 03, 06 and 09.

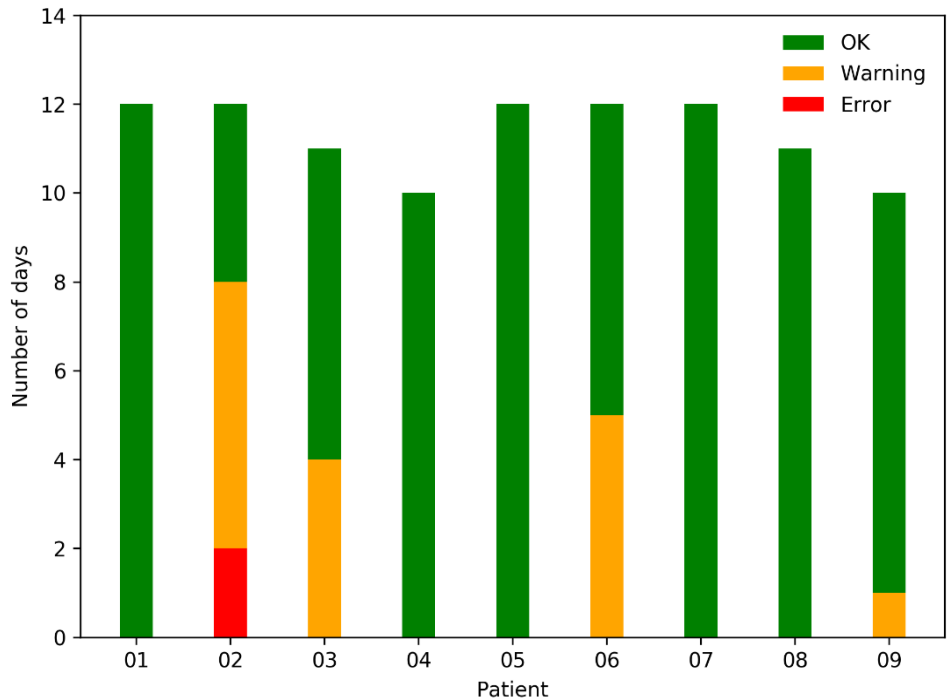

*Figure 9. Daily gross patient error detection with 9 prostate treatment patients. The number of fractions labeled in green indicates the one which has patient variations in tolerance. The number of days labeled in orange indicates the one which has patient variations out of tolerance. And the number of days labeled in red indicates the one which has unaccepted patient variations.*

The EPID-based real-time verification system has been developed in C++ and Python, and we are planning to implement it to the clinical system control computer to automatically receive the EPID frames and perform the comparison. The use of image distance metrics has several advantages. It is much faster than the gamma calculation [37,38] and is robust for a large range of differences between predicted and measured images. There is also a potential to improve the system speed by using hardware such as graphic-based processing units (GPUs) for cine EPID image prediction algorithm which is the most time consuming function of the system.

Although in clinical practice we could use CBCT for patient gross error detection, we did not use CBCT in this study because patient motion can occur post CBCT alignment. However, the combination between EPID-based error detection and other devices such as CBCT can give some additional information or confirmation of the error existence.

There are several potential clinical problems of EPID frame measurement that was neglected in our simulations, e.g. image noise, banding artifacts in single EPID images, sub-mm MLC leaf position variations, and MU difference between measurement and prediction (for pre-computed images). These problems could affect our error detection. There are some studies were done to overcome them, such as aperture edge masking to avoid in-tolerance MLC errors [39]. In addition, the lacking of input samples for error classification can also be an issue since we need a large size of error and no-error samples to cover all types of patient variations and provide a precise classification, especially for the daily changing basis.

# 5 CONCLUSION

The objective of this study is to determine if gross patient miss-alignments or even the incorrect patient can be distinguished from in tolerance patient variations via analysis of the fluence which passes through the beam aperture. The method for gross patient error detection was based on the cine transmission dosimetry (~10 frames per second) of EPID. Different frame-by-frame comparison methods were compared, ROC curves show the best performance belongs to image distance metrics. The results in this study show that the error detection method was able to distinguish gross patient miss-alignment from in-tolerance levels for both 10 H&N and 19 prostate datasets. In addition, for the prostate datasets, the methods utilized were also able to distinguish the incorrect patient errors from the baseline. This study shows that we could prevent rare, but gross and dosimetrically consequential errors which might otherwise be missed in radiation therapy. Furthermore, once gross errors are detectable, then the possibility of detecting smaller non-gross errors becomes possible.

# 6 CONFLICT OF INTEREST

This work is supported by Varian Medical Systems.